\documentclass[utf8]{FrontiersinHarvard} 

\usepackage[onehalfspacing]{setspace}
\usepackage{url,hyperref,lineno,microtype,subcaption}

\def\keyFont{\fontsize{8}{11}\helveticabold }
\def\firstAuthorLast{Polisensky {et~al.}} 
\def\Authors{Emil Polisensky\,$^{1,*}$, Tracy E. Clarke\,$^{1}$, Simona Giacintucci\,$^{1}$ and Wendy Peters\,$^{1}$}
\def\Address{$^{1}$
U.S. Naval Research Laboratory, Code 7213, Washington, DC, USA}
\def\corrAuthor{Emil Polisensky}

\def\corrEmail{emil.j.polisensky.civ@us.navy.mil}

\begin{document}

\onecolumn
\firstpage{1}

\title[Commensal Primary Beam Calibration]{Primary Beam Calibration for Commensal Telescopes Utilizing Offset Optics}

\author[\firstAuthorLast ]{\Authors} 
\address{} 
\correspondance{} 

\extraAuth{}

\maketitle

\begin{abstract}

\section{}

\noindent{\bf Introduction:} Accurate primary beam calibration is essential for precise brightness measurements in radio astronomy. The VLA Low-band Ionosphere and Transient Experiment (VLITE) faces challenges in calibration due to the offset Cassegrain optics used in its commensal observing system. This study aims to develop a novel calibration method to improve accuracy with no impact on the Very Large Array (VLA) primary science observations.

\noindent{\bf Methods:} We used the apparent brightness of standard candles identified in VLITE’s commensal data to develop 1D and 2D primary beam response models. These models accounted for operational changes and asymmetries caused by the subreflector and were validated against holographic methods and compact source light curves.

\noindent{\bf Results:} The models achieved calibration accuracy within 3\% across the field of view, significantly improving the precision of brightness measurements. The results were consistent with holography-derived solutions and performed reliably under different operational conditions.

\noindent{\bf Discussion:} This improved calibration technique expands VLITE’s capabilities for studying active galactic nuclei, transients, and pulsars. It offers a cost-effective alternative to traditional holographic methods, facilitating broader use in commensal observing systems.

\tiny
 \keyFont{ \section{Keywords:} Calibration, Astronomical optics, Astronomical techniques, Flux calibration, Radio telescopes, Radio interferometry} 
\end{abstract}

\section{Introduction}

Radio astronomy is a crucial field in studying astrophysical plasmas through the detection and analysis of radio waves emitted by cosmic ionized gases. These radio emissions offer valuable insights into particle acceleration mechanisms, temperature, density, magnetic fields, and interactions with other forms of matter and energy. Radio astronomy has led to significant breakthroughs, such as the discovery of pulsars, detection of the Milky Way's diffuse synchrotron radiation, and identification of radio galaxies and quasars. It has also been instrumental in studying solar flares, coronal mass ejections, and the interstellar medium. 

In radio astronomy, the fundamental detector element is the antenna. The antenna's sensitivity to radio emissions from the sky is referred to as the primary beam response. In a typical astronomical antenna with a reflecting surface designed to enhance gain in the pointing direction, the primary beam features a prominent peak, known as the main lobe, surrounded by nulls and weaker sidelobes. The primary beam defines the solid angle over which the antenna effectively collects and measures radio emissions from celestial sources.

Radio astronomy uses interferometers to measure Fourier components of the sky brightness distribution by correlating the observed complex voltages between all pairs of antennas in an array. Inverse Fourier transform techniques are employed to construct images of the sky that are convolved with the point-spread function of the array and multiplied by the average primary beam of the antennas. Knowledge of the primary beam is needed to correct for the attenuation and achieve accurate brightness measurements of sources across the field of view. Typically, a singular, symmetric, and non varying model for the primary beam can be assumed and divided out.

\begin{figure}[h!]
\begin{center}
\includegraphics[width=\linewidth]{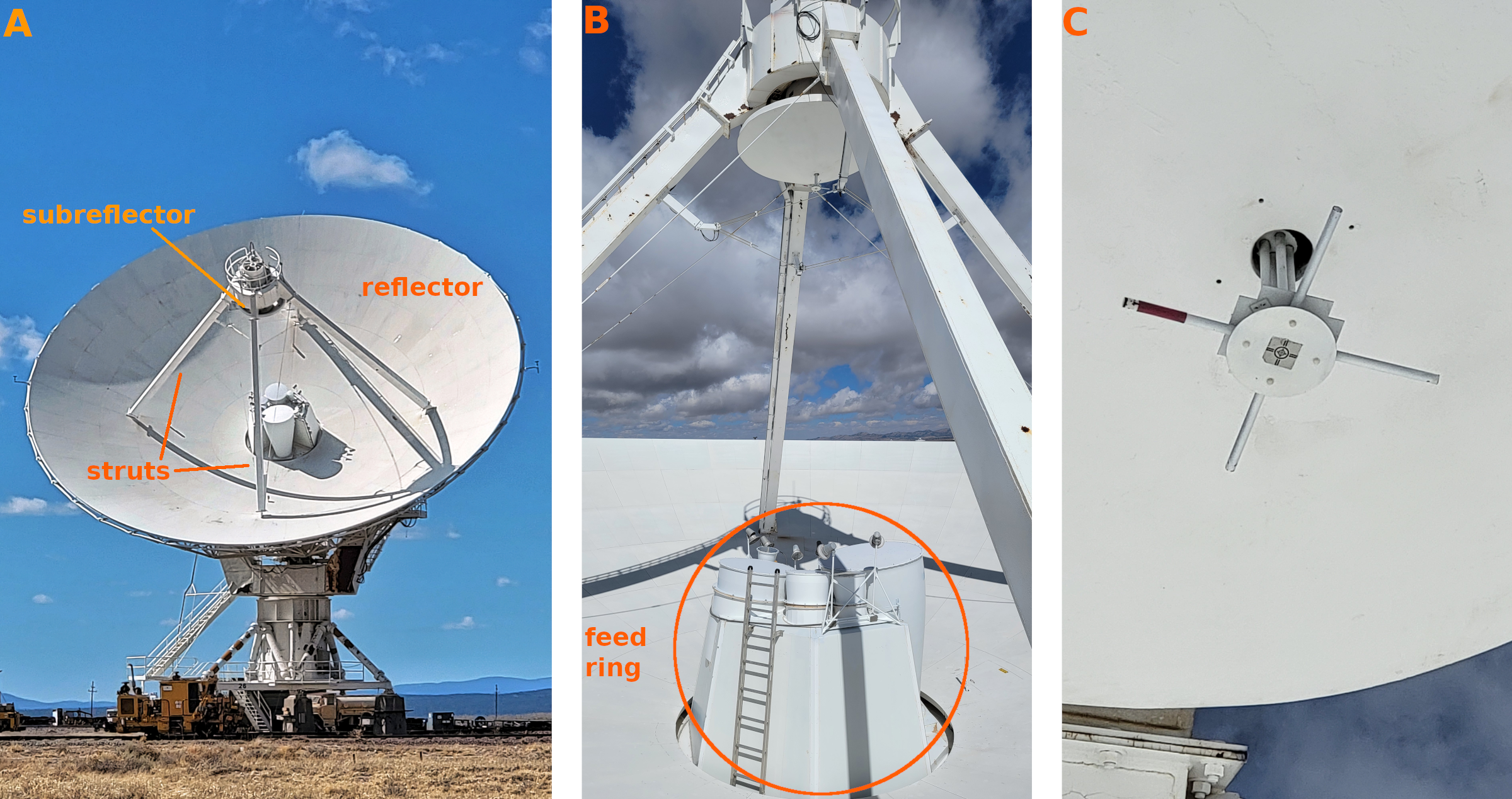}
\end{center}
\caption{Each VLA antenna is an off-axis Cassegrain telescope. \textbf{(A)} The subreflector is held by four struts in front of the parabolic reflector's prime focus. \textbf{(B)} The subreflector is oriented slightly off-axis, rotates, and changes height above the reflector to focus high frequency light to one of eight receivers in the central feed ring at the Cassegrain focus. \textbf{(C)} Close-up of the low frequency P-band receiver, consisting of a pair of crossed dipole antennas protruding from the subreflector and oriented perpendicular to the antenna's pointing axis along the elevation and azimuth axes. The dark band marks the vertically aligned dipole.}\label{fig:vla}
\end{figure}

The National Radio Astronomy Observatory’s Karl G. Jansky Very Large Array (VLA) is a reconfigurable array of 27 antennas situated along the three arms of a rail wye. Each antenna is a 25 m diameter telescope employing Cassegrain optics, as illustrated in Figure~\ref{fig:vla}, with reflecting surfaces specially shaped to improve Cassegrain efficiency. The quasi-parabolic reflector dish redirects incident radio light from the pointing direction toward the prime focus. Four struts hold a rotating quasi-hyperbolic subreflector in front of the prime focus which reflects the light to radio receivers at the Cassegrain focus at the dish center. The frequency range from 1-50 GHz is covered by eight high frequency receivers arranged around a feed ring at the center of the reflector. The name, frequency range and angle in the ring for each of these high frequency receivers is given in Table \ref{table:bands}. A more detailed description of the VLA and receivers is given in \cite{perley2011}.

\begin{table}[h!]
\caption{VLA Cassegrain Focus Receivers \label{table:bands}}
\centering
\begin{tabular}{lcccccccc}
\hline
\hline 
Band Name & L & S & C & X & Ku & K & Ka & Q \\
Frequency Range (GHz) & 1-2 & 2-4 & 4-8 & 8-12 & 12-18 & 18-26 & 26-40 & 40-50 \\
Offset Angle$^*$ ($^\circ$) & -174.1 & 11.6 & 75.2 & 113.7 & -42.4 & -64.1 & -106.9 & -85.5 \\ [1ex]
\hline
\multicolumn{9}{l}{$^*$ Angles are measured counterclockwise from the zenith direction looking into the dish.}\\
\end{tabular}
\end{table}

Observations in the 224-480 MHz range (P-band) are enabled by a receiver consisting of a pair of linearly polarized dipole antennas located near the prime focus of the reflector. The dipoles are mounted to the antenna structure, protrude from the subreflector along the pointing axis of the reflector and are aligned perpendicular to this axis. The subreflector acts as a ground plane for the dipoles but also creates complications for the P-band primary beam pattern.

The first complication is that the P-band receiver is never precisely at the prime focus but is instead positioned significantly in front of it. This offset causes the primary beam to become defocused, resulting in a broad plateau surrounding the central Gaussian-shaped main lobe \citep{evla195}. Additionally, while the P-band dipoles remain fixed, the subreflector adjusts its distance from the reflector to focus light correctly on the desired feed height in the central ring. These subreflector adjustments alter both the shape and strength of features within the P-band primary beam, as the dipoles are positioned at a non-standard distance from the ground plane, deviating from the canonical quarter-wavelength separation.

\begin{figure}[h!]
\begin{center}
\includegraphics[width=\linewidth]{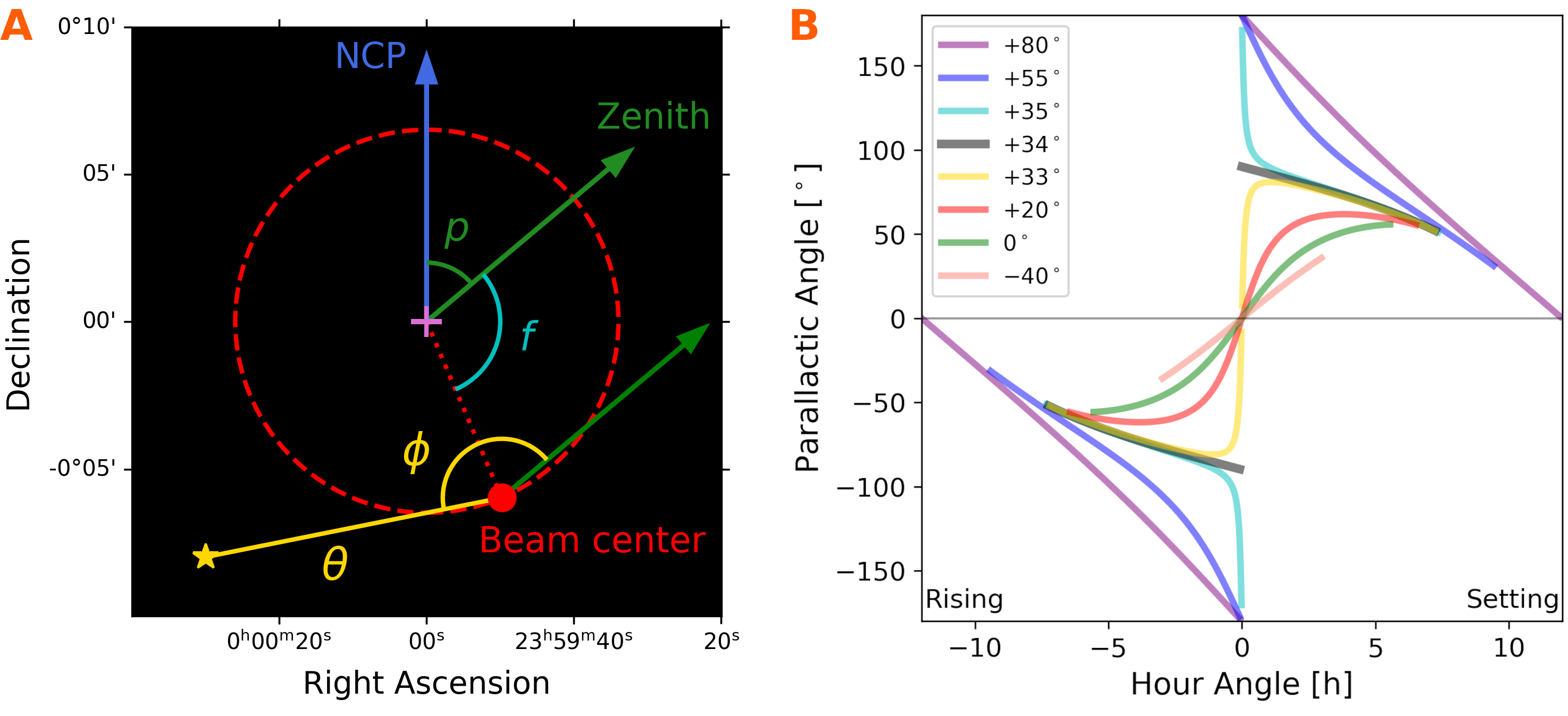}
\end{center}
\caption{\textbf{(A)} Illustration of the offset, rotating VLA P-band primary beam for a pointing direction of ($0^\circ$,$0^\circ$). The beam center is offset by $6.5^\prime$ (dotted line) in a direction that depends on the parallactic angle $p$ and rotation angle of the subreflector $f$. As $p$ changes during the observation, the beam center rotates around the dashed circle. The coordinates $\theta$ and $\phi$ are described in Section \ref{sec:vlitedata}. \textbf{(B)} Variation in the parallactic angle for sources observed from the latitude of the VLA. Source tracks for several declinations are shown. Sources rise above the horizon at negative hour angle, pass the meridian at 0 hour angle, and set below the horizon at positive hour angle. A minimum observing elevation of $5^\circ$ is assumed.}\label{fig:coords}
\end{figure}

The subreflector is also tilted relative to the pointing axis to focus high frequency light to the receivers on the feed ring. The tilt offsets the center of the P-band beam from the pointing axis by $6.5^\prime$ (K. Sowinski, private communication) in the direction of the Cassegrain feed in use. The direction of the offset is stationary in the horizon coordinate system of the alt-azimuth mounts used by the VLA antennas. However, in the equatorial coordinates used in imaging, the offset rotates about the pointing direction as the parallactic angle between the zenith direction and the north celestial pole direction changes during an observation. This is illustrated in Figure \ref{fig:coords}. The tilted subreflector also presents an asymmetric ground screen to the P-band dipoles that extends longer in the radial direction away from the high frequency feed in use, creating an asymmetry in the beam pattern.

A novel approach in radio astronomy is commensal observations, where an independent data collection system utilizes the existing infrastructure of an instrument. The VLA Low-band Ionosphere and Transient Experiment (VLITE) is a pioneering instrument in this domain. VLITE records data using the P-band feed near the prime focus while the VLA records data using the high-frequency feeds at the Cassegrain focus. Operating on a subset of antennas and the available bandwidth, VLITE has been recording data during nearly all routine operations on the VLA for the last ten years.

The VLA’s offset Cassegrain optics pose challenges for primary beam calibration in VLITE. The VLITE primary beam is asymmetric and varies in shape depending on the high-frequency receiver in use. It is also offset from the pointing axis and rotates around it, with the rotation pattern determined by the receiver and parallactic angles of the primary observations. These factors result in a unique attenuation pattern which must be individually computed for each VLITE image.

This paper is organized as follows: in Section 2 we describe our technique to map the VLITE primary beam and compute the unique calibration function for each VLITE image. We present the resulting primary beam maps in Section 3, compare to the map obtained from traditional methods and test the accuracy of our calibration. We conclude in Section 4 with a discussion of the impact primary beam calibration has on astrophysical plasma research with VLITE.

\section{Methods}

A traditional approach for measuring the primary beam is holography, where a bright, unresolved source is observed with a pair of antennas acting as a two element interferometer \citep{scott1977}. One antenna tracks the source while the other points in an offset direction, allowing for the measurement of the amplitude and phase of its voltage pattern. By taking multiple measurements at various offsets, a comprehensive map of the primary beam can be constructed and analyzed. This method offers the advantage of detailed beam mapping for each antenna in the array and can identify reflector deformations and optical misalignments, which can then be corrected to enhance antenna performance \citep{mayer1983}. 

The holographic method, however, requires significant observing time, which conflicts with VLITE’s goal of not interfering with the VLA’s primary science observations. Instead, we use the apparent brightness of a population of standard candles identified within the commensal data. By comparing the apparent brightness of the ensemble of sources scattered across the field of view with their known values, we can statistically determine the primary beam attenuation. With additional data on the sources' distances from the beam center and their position angles, we can map and model the two-dimensional primary beam response for each subreflector position.

We opted to use source brightness rather than flux density for two main reasons. First, source brightness derived from fitting 2D elliptical Gaussians to pixel clusters is not affected by the uncertainties in fitted sizes compared to flux density, which is influenced by both signal-to-noise ratio (S/N) and the fitted sizes \citep{condon1997}. Second, fitted Gaussians are often biased to larger sizes due to the inclusion of noise pixels around the island perimeter \citep{hales2012, hop2015}. This bias is particularly significant for VLITE sources due to the highly non-Gaussian noise in VLITE images. Our tests showed that brightness ratios produced a consistent attenuation pattern across different S/N bins, whereas flux density ratios did not. However, using brightness requires a sample of unresolved sources, as the apparent brightness of extended sources can be affected by the absence of short baselines in the array.

Our objective is to model the VLITE primary beam with sufficient accuracy to minimize beam calibration uncertainty in total intensity images across a large field of view. Although this method relies on total intensity images rather than measuring the complex voltage patterns of each dipole in the P-band feed, it effectively meets our requirement for achieving accurate beam calibration, even though it does not allow us to assess the quality of the VLA optics.

\subsection{Catalog Selection}\label{sec:cat}

The VLA is a reconfigurable array with four main configurations, labeled A through D. One complete cycle through all four configurations takes approximately 16 months. VLITE achieves its highest sensitivity and resolution when the VLA antennas are in the A configuration, which features the longest baselines of $\sim 35$ km. Conversely, sensitivity and resolution are lowest in the D configuration, where the antenna distribution is most compact with maximum baselines $\sim 1$ km. The size of the field of view imaged by the standard VLITE imaging pipeline varies with configuration due to the computational constraints imposed by the higher number of pixels required for increased resolution. Images from the D configuration extend over a diameter of $8^\circ$, while those from the A configuration are limited to $2^\circ$.

To achieve accurate VLITE primary beam calibration over the largest possible solid angle, we require a catalog of standard candles that is both modest in resolution and extensive in coverage. The catalog should also be at a frequency similar to that of VLITE to minimize uncertainties arising from differences in source spectra. The Westerbork Northern Sky Survey \citep[WENSS,][]{wenss} meets these requirements effectively. WENSS covers 10,000 square degrees (declination $> +30^\circ$) at a frequency of 325 MHz. Its resolution of approximately $54^{\prime\prime}$ closes matches the $32^{\prime\prime} - 66^{\prime\prime}$ resolution of VLITE images in the C configuration. Its detection threshold of $\sim$ 18 mJy is deep enough to provide matches for over 97\% of VLITE C configuration sources.

To identify unresolved sources, we use a well-established method \citep[e.g.,][]{tgss,degas2018,frail2024} based on a compactness metric derived from the flux density-to-brightness ratio $R$ and the S/N. This approach is necessary because measurement errors in flux density and brightness are correlated with S/N. Various methods can determine a smoothly varying function in $R$-S/N space that statistically differentiates unresolved from extended sources. For the WENSS survey, this separation was accomplished using Monte Carlo simulations \citep[][Eq. 10 with $C = 3.2$]{wenss}. The compactness metric is the ratio of $R$ from the fitted equation to the measured value. Sources with a compactness $> 1$ are $\sim 97\%$ likely to be unresolved.

We define a high-quality sample of unresolved sources in WENSS by selecting single-component sources with compactness $> 1$ and excluding those where the source-finding algorithm failed or where non-zero major and minor axes indicate likely resolved sources. This process yields a catalog of 167,647 WENSS sources.

Given the sensitivity of the WENSS survey, our source sample predominantly consists of active galactic nuclei \citep[AGN,][]{nvss}. However, because WENSS observations were conducted between 1991 and 1997 — about 20-30 years prior to VLITE observations — there are concerns about using these sources as standard candles. Over this period, sources may have experienced brightness variations due to intrinsic factors such as changes in accretion physics or jet orientation, as well as extrinsic factors like scintillations from electron density irregularities in the interstellar medium. These variations can both increase and decrease source brightness. We thus expect that a sufficiently large sample will statistically balance out these variations and provide an accurate measure of the primary beam, albeit with greater scatter than expected from measurement error alone.

\subsection{VLITE data}\label{sec:vlitedata}

VLITE began science operations in November 2014, running commensally with 10 VLA antennas during nearly all regular VLA observations at GHz frequencies. In July 2017, VLITE expanded to 18 antennas to significantly enhance its sensitivity and imaging fidelity. For this study, we use only data collected after this expansion. VLITE images produced by the standard imaging pipeline have an effective bandwidth of approximately 40 MHz centered at 340 MHz due to design limitations and persistent radio frequency interference (RFI).

The VLITE Database Pipeline \citep[VDP,][]{vdp} processes images by cataloging sources with S/N $> 5$. Sources in images that pass quality controls are associated with previous detections to track multiple observations of unique sources. Separate catalogs are maintained for different VLA array configurations, which affect image resolution, and these catalogs are matched to other radio catalogs with similar angular resolutions. In particular, sources in C configuration images are automatically matched to the WENSS catalog. C configuration images cover a field of view $6^\circ$ in diameter, with sources at the corners of the imaged field of view extending up to $\sim 4^\circ$ from the beam center.

Compactness metrics are calculated for VLITE sources in a similar manner as for WENSS but statistical methods are used rather than injection simulations \citep[e.g.][]{williams13}. Compactness analysis is performed separately for each array configuration within each 16-month cycle to account for variations in the VLITE antenna distribution and source fitting bias. The VDP stores compactness data, source finding products, and catalog matches in a Structured Query Language (SQL) database for efficient access and selection.

We select VLITE source detections that match the WENSS unresolved source sample based on the following criteria: sources must be fit as single-component Gaussians, have compactness $> 1$, and have been modeled by the CLEAN algorithm during the imaging step. Additionally, the fractional uncertainty in the VLITE-WENSS brightness ratio must be $< 10\%$, derived from the propagated uncertainties of the VLITE and WENSS brightness measurements.

Longer-duration observations improve sensitivity and measurement precision but also cause sources to sample a smeared beam due to the changing parallactic angle. To assess the impact of including long-duration images, we compared our fitted models with those derived from images where the change in parallactic angle was kept below $10^\circ$, corresponding to maximum durations of 16,000 seconds. We found that beam smearing effects were most pronounced at larger distances, where asymmetries in the beam shape are greater, though they remained under $2\%$ within $2^\circ$ and under $3\%$ within $3^\circ$. However, this restriction reduces the number of points sampling the beam by a factor of 2-3. To strike a balance between sensitivity and beam sampling quality, we restrict our sample to images with durations below $20,000$ seconds. 

Shorter-duration images, while providing better sampling of the unsmeared beam due to minimal parallactic angle change, tend to be less sensitive and more affected by noise. Our testing indicates that applying a minimum parallactic angle change $> 4^\circ$ yields a higher quality sample than imposing a minimum image duration. With these criteria, we obtain 80,595 VLITE detections of 17,218 unique WENSS sources across 2,359 images.

We define a coordinate system where each source's position in the primary beam is specified by the angles $\theta$ and $\phi$. The angle $\theta$ measures the angular distance from the beam center, while $\phi$ measures the polar angle. In this system, the subreflector is aligned vertically toward the zenith direction ($0^\circ$ feed angle). Consequently, $\phi$ measures the angle from the zenith vector in the image plane. These coordinates are illustrated in Figure \ref{fig:coords}A for a source marked by the star symbol.

To determine the coordinates for each source, a single parallactic angle for the image is first computed as the average of the angles at the beginning and end of the observation. The location of the beam center is calculated from this parallactic angle, the subreflector rotation angle of the high frequency feed used during the observation, and the offset magnitude of $6.5^\prime$. The coordinate $\theta$ is determined as the angular separation between the source and the beam center. The zenith vector is found by adding the parallactic angle to the north celestial pole vector at the position of the beam center. Finally, the coordinate $\phi$ is computed as the angle to the source east of the zenith vector.

Using the coordinates $\theta$ and $\phi$, the VLITE-to-WENSS brightness ratio, and information about the high frequency feed employed during the observation, we can fit models designed to account for the defocusing and asymmetries observed in the VLITE beam pattern.

\subsection{Data Combination}

The phase centers of the high-frequency feeds are positioned at different heights above the reflector. The subreflector adjusts its distance from the reflector to focus light on these phase centers. Additionally, the orientation of the asymmetric subreflector with respect to the P-band dipoles varies with the feed ring angle of the high-frequency feed. 

However, since we are modeling the total intensity primary beam profiles, any dependency in the responses of the individual, linear polarized dipoles due to changes in the subreflector orientation should average out. By accounting for the feed ring position angle and calculating the polar angle $\phi$ in a frame where the subreflector is aligned with the zenith, we can combine data from subreflector positions at equivalent distances.

The L and S-band feeds, being at distinct heights, are modeled separately as the "L" and "S" VLITE primary beams. In contrast, the Ku, K, Ka, and Q-band feeds share the same height and are combined into the "KQ" primary beam. Although the C and X-bands differ by about 4\% of a VLITE wavelength, testing showed no significant benefit from fitting them separately, so they are grouped together as the "CX" beam. Combining data from feeds at the same or similar subreflector distances enhances the number of samples and the robustness of the model fits. Specifically, the S beam has about 9,000 samples, the L beam has about 42,000, and both the CX and KQ beams have approximately 14,000 samples each.

\subsection{1D Primary Beam Model}

We first describe our model for the primary beam as a function of radial distance from the beam center, $\theta$. All model fits use data weighted by the inverse of the brightness ratio variance.

The main lobe is modeled as a Gaussian:
\begin{equation}\label{eq:gauss}
   P(\theta) = a_0 \exp(-\theta^2/2\sigma^2)
\end{equation}

The parameter $\sigma$ defines the width of the main lobe and varies with the distance of the subreflector from the prime focus. The central value $a_0$ represents the beam normalization, determined by flux scale calibration differences between WENSS and VLITE and the average spectral scaling between 325 MHz and 340 MHz. Since $a_0$ is independent of subreflector position, it can be determined separately from $\sigma$. This approach is more robust than treating $a_0$ as a free parameter for each subreflector position.

\begin{figure}[h!]
\begin{center}
\includegraphics[width=4in]{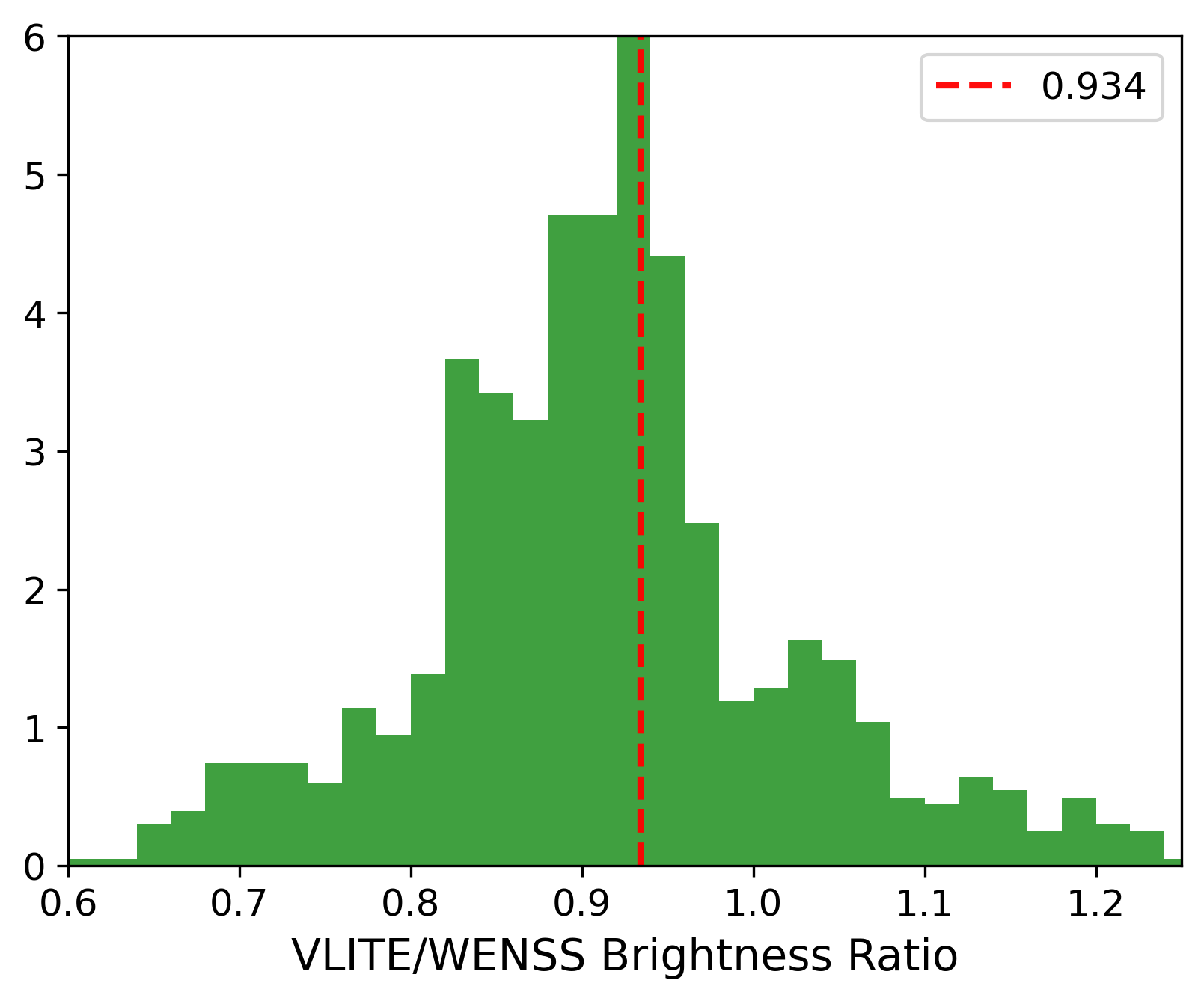}
\end{center}
\caption{Histogram of the brightness ratios of unresolved WENSS sources detected within $8^\prime$ of the VLITE primary beam center. The weighted average of 0.934 sets the normalization of the primary beam.}\label{fig:center}
\end{figure}

To determine $a_0$, we select sources with $\theta < 8^\prime$. This broader cut is necessary to ensure a sufficient sample, as many sources are observed near the pointing center, which is $6.5^\prime$ from the beam center. The primary beam varies by less than 1\% within the inner $8^\prime$, making this selection appropriate. We also select strongly detected sources with S/N $> 30$ to obtain a robust estimate of $a_0$. Figure \ref{fig:center} shows the brightness ratio distribution of 1,038 detections from 198 unique WENSS sources meeting these criteria. The weighted mean value of $a_0$ is 0.934, which falls within the bin with the highest number of sources. However, the distribution is unexpectedly non-Gaussian and warrants further examination.

The distribution reveals a significant number of outliers with brightness ratios $20–30\%$ from the mean, likely due to highly variable blazars. Around the average, the distribution shows a notable skew toward smaller values. This trend is not limited to central sources but is ubiquitous across distance. We speculate that, despite selecting compact sources, some WENSS sources are extended and partially resolved by VLITE's slightly higher resolution. Supporting this hypothesis, detections in the 2017A and 2020A observing semesters — when VLITE had lower resolution — show less skew than those in the 2018B and 2021A semesters when the resolution was higher.

Since $a_0$ determines the strength of the defocused outer plateau relative to the main lobe, our calibration solution is highly sensitive to $a_0$. Testing in Section \ref{sec:calibtests} showed that even a small variation of 0.01 in $a_0$ causes noticeable systematic effects, reinforcing confidence in our adopted value of 0.934, despite the broader spread seen in Figure \ref{fig:center}.

To determine $\sigma$, we perform a weighted fit of Equation \ref{eq:gauss} using data points where $\theta < 1.3^\circ$, excluding the central sources. This selection helps to minimize contamination from the defocused plateau of the outer beam, which could otherwise bias $\sigma$ towards larger values.

We fit the intricate shape of the outer beam with a penalized cubic spline using the ALGLIB\footnote{www.alglib.net} library. Penalized splines are effective for handling complexity in noisy data and offer ease of tuning, with only two free parameters: the number of spline nodes $M$ and the regularization coefficient $\rho$. The parameter $M$ must be sufficiently large to ensure flexibility but not so large as to impair performance, while $\rho$ controls the degree of smoothing. Through experimentation, we determined that $M=200$ and $\rho=4$ produced smooth fits that accurately captured details of the beam shape without overfitting. We also tested the method of locally weighted scatterplot smoothing \citep[LOWESS,][]{lowess}, which is discussed further in Section \ref{sec:calibtests}.

We define $\theta_X$ as the cross-over point where the fitted Gaussian value falls below the spline, marking the transition from the inner main lobe to the outer plateau model. The cross-over point is $\sim 1^\circ$ for all fits. Since the true beam is expected to vary smoothly with a continuously varying derivative, we use a smoothing function, $S(\theta)$, to mitigate the jump discontinuity in the slope at $\theta_X$:
\begin{equation}
   S(\theta) = \frac{1}{2} + \frac{1}{2} \tanh((\theta - \theta_X)/b)
\end{equation}
The parameter $b$ controls the width of the smoothing function. We set $b=0.15$ to limit the smoothing effect to a small range, ensuring a smooth transition between the Gaussian and spline models while minimizing disruptions to the fitted values.

The final 1D beam model with the smoothing applied is calculated as:
\begin{equation}
   P(\theta) = (1-S(\theta)) P_{Gauss}(\theta) + S(\theta) P_{spline}(\theta)
\end{equation}

\begin{figure}[h!]
\begin{center}
\includegraphics[width=\linewidth]{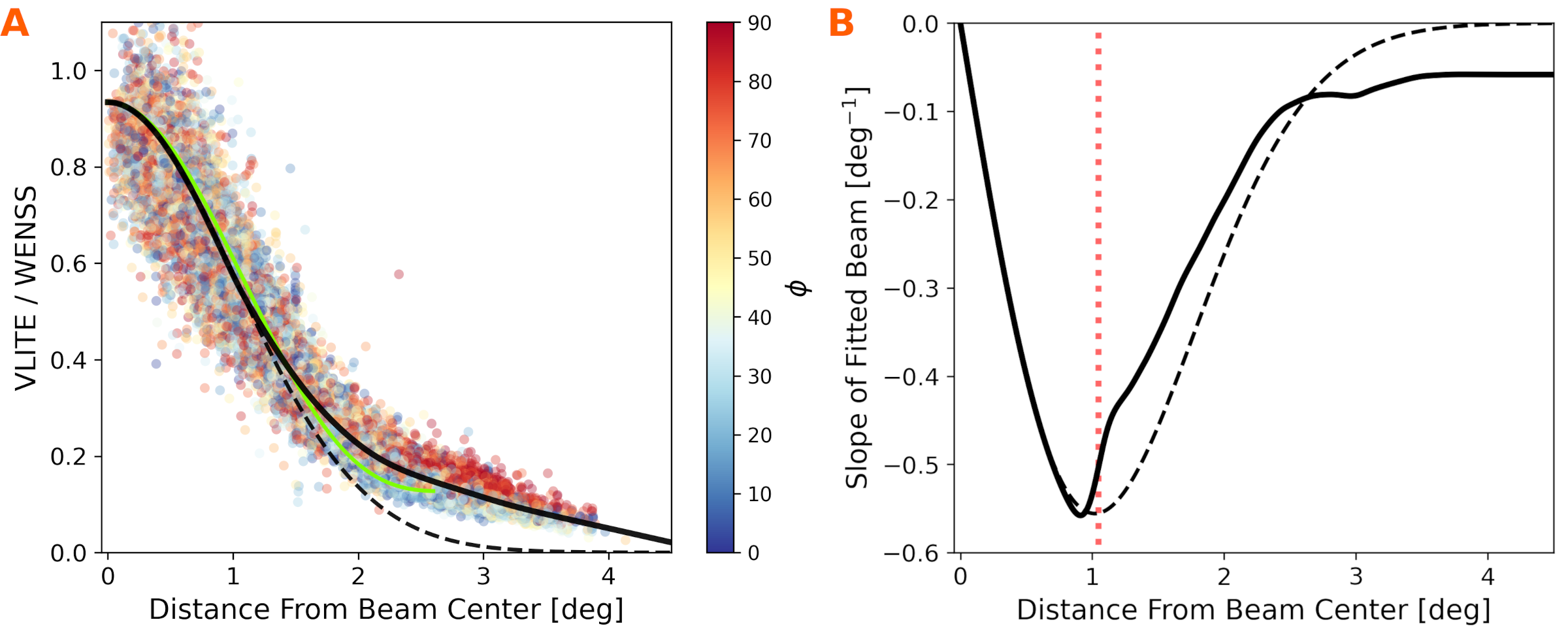}
\end{center}
\caption{\textbf{(A)} Brightness ratios with distance from beam center ($\theta$) for sources observed in the L-band subreflector position. Data points are colored by their polar angle $\phi$. Black line shows the fitted 1D beam model (Gaussian main lobe $+$ outer spline). Dashed line shows the Gaussian fit to the main lobe. Green line shows the 344 MHz VLA primary beam model determined from holography. \textbf{(B)} Derivative of the fitted beam model. Dotted vertical line shows the cross-over angle $\theta_X$ between the inner Gaussian and the outer spline models. A hyperbolic tangent function smooths the transition of the slope around $\theta_X$.}\label{fig:fitbeam}
\end{figure}

Figure~\ref{fig:fitbeam}A illustrates the 1D model fitted to the L-band subreflector position, while Figure~\ref{fig:fitbeam}B presents the slope of the fit. The model transitions from the Gaussian shape of the main lobe to the outer spline at $\theta_X = 1.05^\circ$. The smoothing functions introduces a slight steeping of the Gaussian, followed by a flattening just before $\theta_X$. The plateau is flatter than the Gaussian for $\theta \sim 1 - 2.5^\circ$ and becomes steeper beyond $2.5^\circ$.

The data points in Figure~\ref{fig:fitbeam}A are color-coded by polar angle $\phi$. The plot reveals a clear asymmetry in the outer beam ($\theta > 1.8^\circ$) that the 1D model alone cannot capture.

\subsection{2D Primary Beam Model}

To address beam asymmetries, we generate a 2D primary beam model by applying the 1D model to data binned by polar angle $\phi$. We exploit the reflection symmetry about the $0^\circ - 180^\circ$ and $90^\circ - 270^\circ$ axes to fold the data into the first quadrant: $0^\circ < \phi < 90^\circ$, and increase the number of data points in each bin.

We sample $\phi$ in $1^\circ$ increments and select data points within $\pm10^\circ$ of $\phi$. The $10^\circ$ range ensures robust fits, particularly for the $2-4$ GHz S-band beam, which has the fewest data points. We fit the 1D model to the data within each $\phi$ bin and tabulate the fitted values in $1^{\prime\prime}$ steps in $\theta$. A Savitzky-Golay filter, with a window length of 19 and a 3rd-order polynomial, is applied to smooth the values in $\phi$ for each distance step in $\theta$. For the final 2D beam model, we retain the tabulated 1D fits for each $5^\circ$ step in $\phi$. The model values are normalized by $a_0$ and saved in text files.

\subsection{Image Primary Beam Calibration}\label{sec:imagecalib}

This section describes how the unique beam attenuation function for each VLITE image is computed from the tabulated primary beam models.

Each VLITE image consists of one or more observation "blocks" during which the position on the sky is tracked continuously. The VLITE imaging pipeline generates data products that detail the number of blocks, their start and stop times, and the number of recorded visibilities. The continuous smear of the rotating primary beam within each block is approximated by discrete sampling. At a minimum, each block is sampled at its start and end. If a block's duration exceeds a specified "smear time," additional samples are taken. The number of extra samples is determined by the integer ratio of the block duration to the smear time, with sample times evenly spaced. We adopt a smear time of 900 seconds and evaluate the accuracy of this sampling in Section \ref{sec:calibtests}.

At each sample time, the parallatic angle is computed and used to determine $\theta$ and $\phi$ for all pixels. These angles are then converted to array indices using the tabulated model sampling values ($1^{\prime\prime}$ for $\theta$ and $5^\circ$ for $\phi$), allowing for straightforward lookup of the primary beam value for each pixel.

\begin{figure}[h!]
\begin{center}
\includegraphics[width=\linewidth]{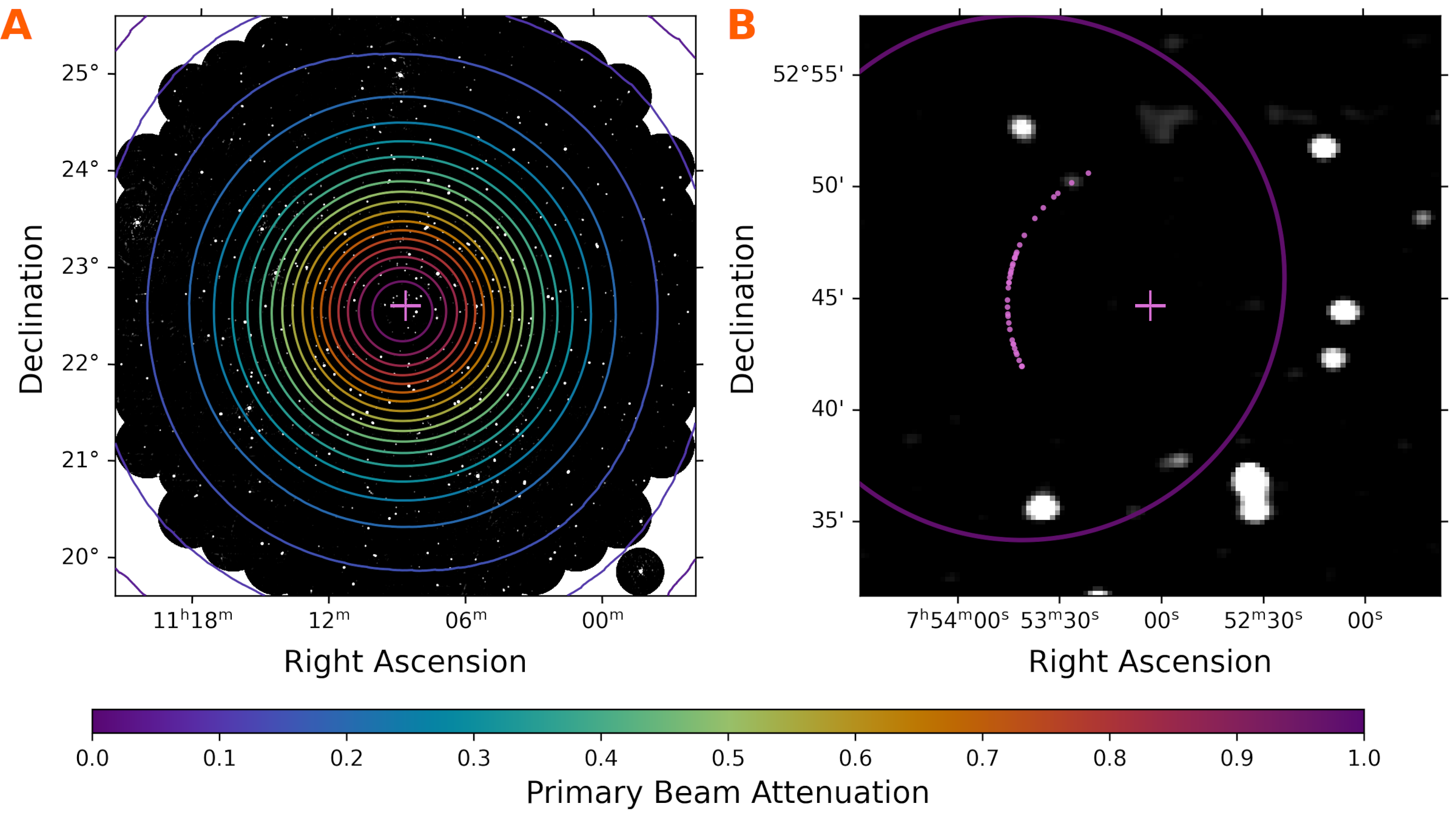}
\end{center}
\caption{Contours of primary beam attenuation overlain on VLITE images. Crossed lines indicate the pointing center for each image. \textbf{(A)} A $6^\circ$ wide C configuration image. The Gaussian main lobe and the asymmetric, extended plateau of the outer beam are evident. \textbf{(B)} The central region of a C configuration image with the $98\%$ contour shown. Points represent the beam center sampling as it rotated with the parallactic angle on an arc $6.5^\prime$ from the pointing center. The arc shows numerous gaps when the field was either not observed or data were corrupted by RFI.}\label{fig:contours}
\end{figure}

Each block is assigned a weight based on the ratio of its visibilities to the total number of visibilities in the observation. This weight is distributed among the beam samples within the block. The primary beam value for each pixel at each sample time is then weighted and summed to create an image of the primary beam attenuation across the field of view. Efficient computations are performed using NumPy array methods. While long-duration observations with many blocks can take several minutes, most VLITE images have their beam attenuation computed in $< 30$ seconds. Figure~\ref{fig:contours}A illustrates the calculated primary beam attenuation as contours on a VLITE image, while Figure~\ref{fig:contours}B details the block sampling for the beam rotation around the image center.

Special considerations are required for images from the VLITE Commensal Sky Survey \citep[VCSS,][]{VCSSmemo}, a low-frequency survey conducted alongside the 3 GHz VLA Sky Survey \citep[VLASS,][]{vlass}. VLASS is a large, multi-epoch sky survey conducted at S-band with the VLA in the B configuation. VLASS scans the sky at declinations $> -40^\circ$ using an "on-the-fly" mode, where antennas continuously slew in right ascension along strips of constant declination. VLITE data are collected as antennas move through an angular distance of $1.5^{\circ}$, corresponding to $28-58$ seconds of observation time for the declination-dependent VLASS slew rates.

\begin{figure}[h!]
\begin{center}
\includegraphics[width=\linewidth]{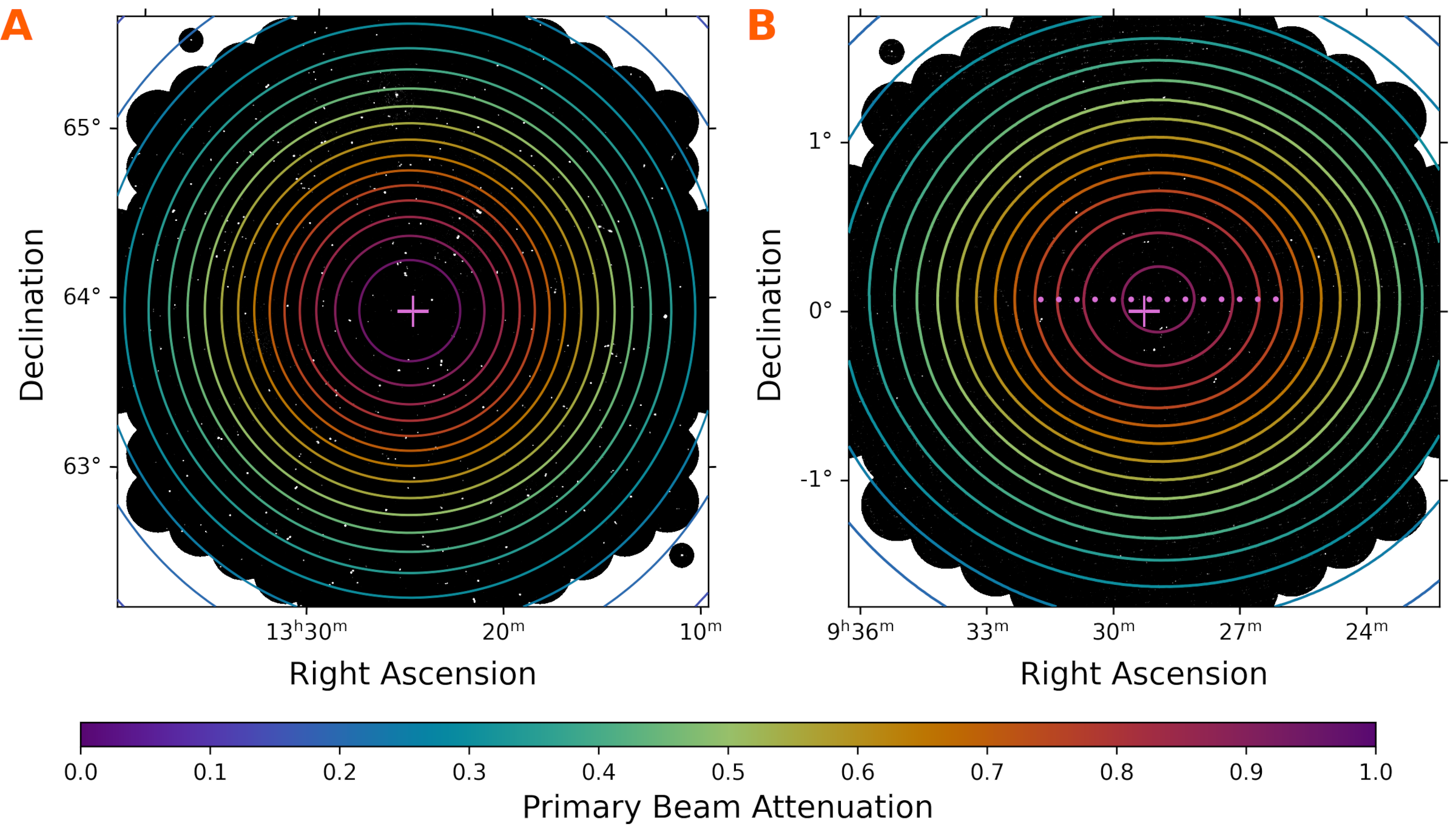}
\end{center}
\caption{Contours of primary beam attenuation overlain on $3.5^\circ$ wide B configuration images. \textbf{(A)} A standard VLITE image. \textbf{(B)} A VCSS snapshot image. The on-the-fly observing mode smears the VLITE primary beam by $1.5^\circ$ in right ascension, producing an elongated attenuation pattern offset from the pointing center. Points indicate the beam center sampling during the smear.}\label{fig:contoursVCSS}
\end{figure}

The data is then correlated and imaged with the midpoint of the antenna motion used as the phase center. This results in short snapshot images with an elongated primary beam response. Calculating the primary beam attenuation for VCSS snapshots requires accounting for the lateral motion in right ascension. The data are treated as a single observing block and sampled at 14 intervals, equally spaced in time, across the $1.5^{\circ}$ motion in right ascension. Figure~\ref{fig:contoursVCSS} compares the primary beam attenuation of a VCSS snapshot with a typical B configuation VLITE image.

\section{Results}

In this section we show the fitted primary beam models, compare to the beam model from holography, and test the accuracy of our image calibration method.

\subsection{Primary Beam Models}

\begin{figure}[h!]
\begin{center}
\includegraphics[width=\linewidth]{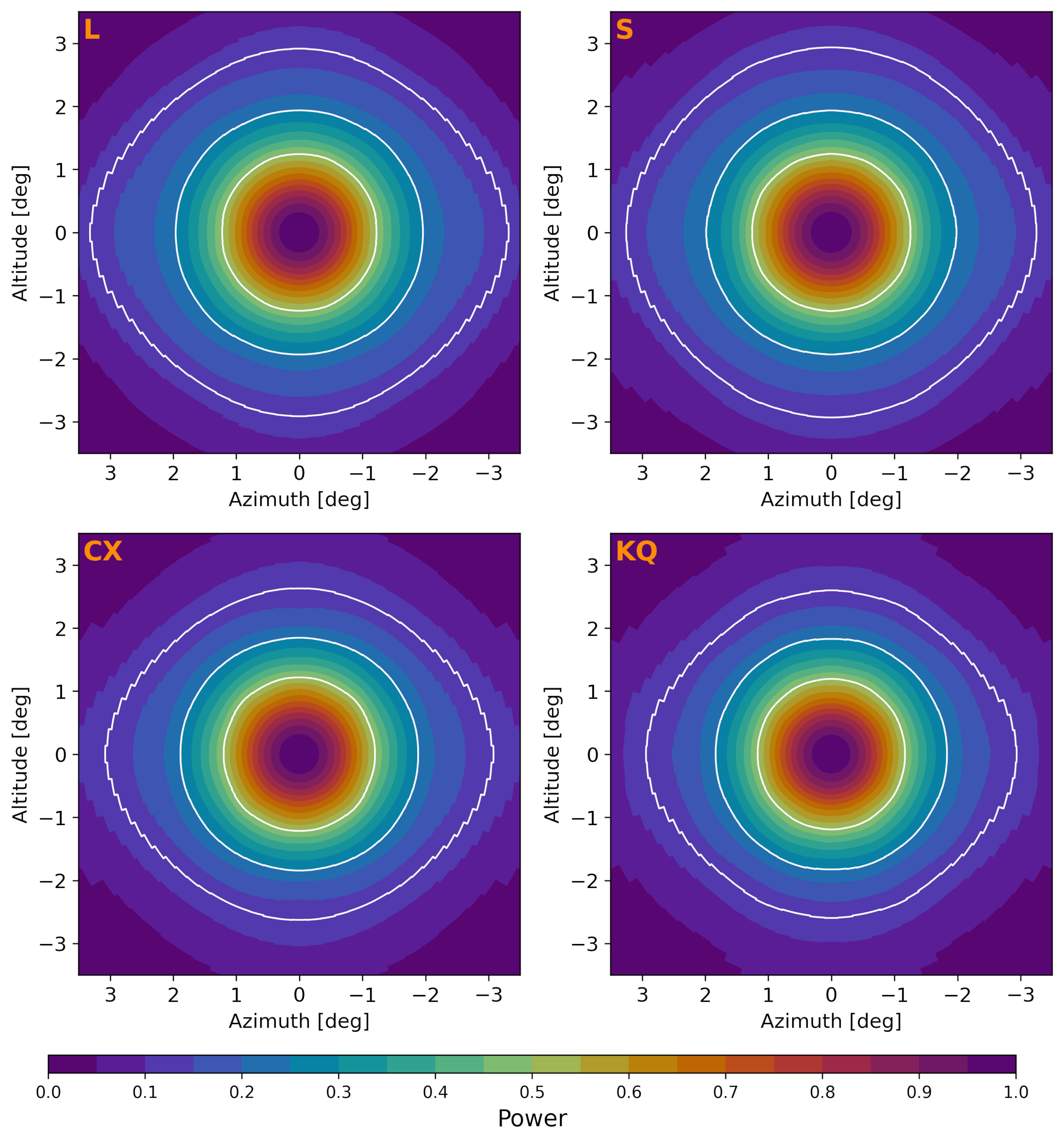}
\end{center}
\caption{The four VLITE primary beams derived from 2D model fitting the source data. Shown are the subreflector positions for the L-band, S-band, C and X-bands (CX), and the Ka, K, Ku, and Q-band feeds (KQ). The subreflector positions are arranged with increasing proximity to the prime focus from left to right and top to bottom. White contours represent the half-power, quarter-power, and one-eighth-power levels. The beams exhibit notable asymmetry beyond $\sim 2^\circ$ from center and a narrowing as the subreflector moves away from the P-band dipoles and closer to the prime focus.}\label{fig:maps}
\end{figure}

Figure~\ref{fig:maps} displays the 2D models fitted to the data for the four VLITE primary beams. The white contours represent half, quarter, and one-eighth power levels, highlighting the size and asymmetries of the beams. The beams are narrower in altitude due to the tilt of the subreflector, which presents an elongated ground plane to the P-band dipoles in this direction. Additionally, the beam widens as the subreflector is positioned closer to the dipoles.

The struts are aligned horizontally and vertically along the azimuth and altitude axes in these plots. Comparing the patterns in the outer beams along diagonal versus horizontal and vertical directions reveals asymmetries likely attributable to these struts.

\subsection{Comparison to the VLA Model}

During dedicated P-band observations with the VLA, the subreflector is positioned as close to the prime focus as possible, maximizing its distance from the P-band dipoles. This setup differs from its position during high-frequency observations. The primary beam has been mapped in this configuration using the holographic technique and modeled with symmetric eighth-degree polynomials across the full P-band frequency range \citep{evla195}. These models achieve about 1.5\% accuracy for distances from center up to $2.6^\circ$. The green line in Figure~\ref{fig:fitbeam}A shows the holographic model (scaled by $a_0$) in comparison to the 1D VLITE data for the L-band position.

\begin{figure}[h!]
\begin{center}
\includegraphics[width=\linewidth]{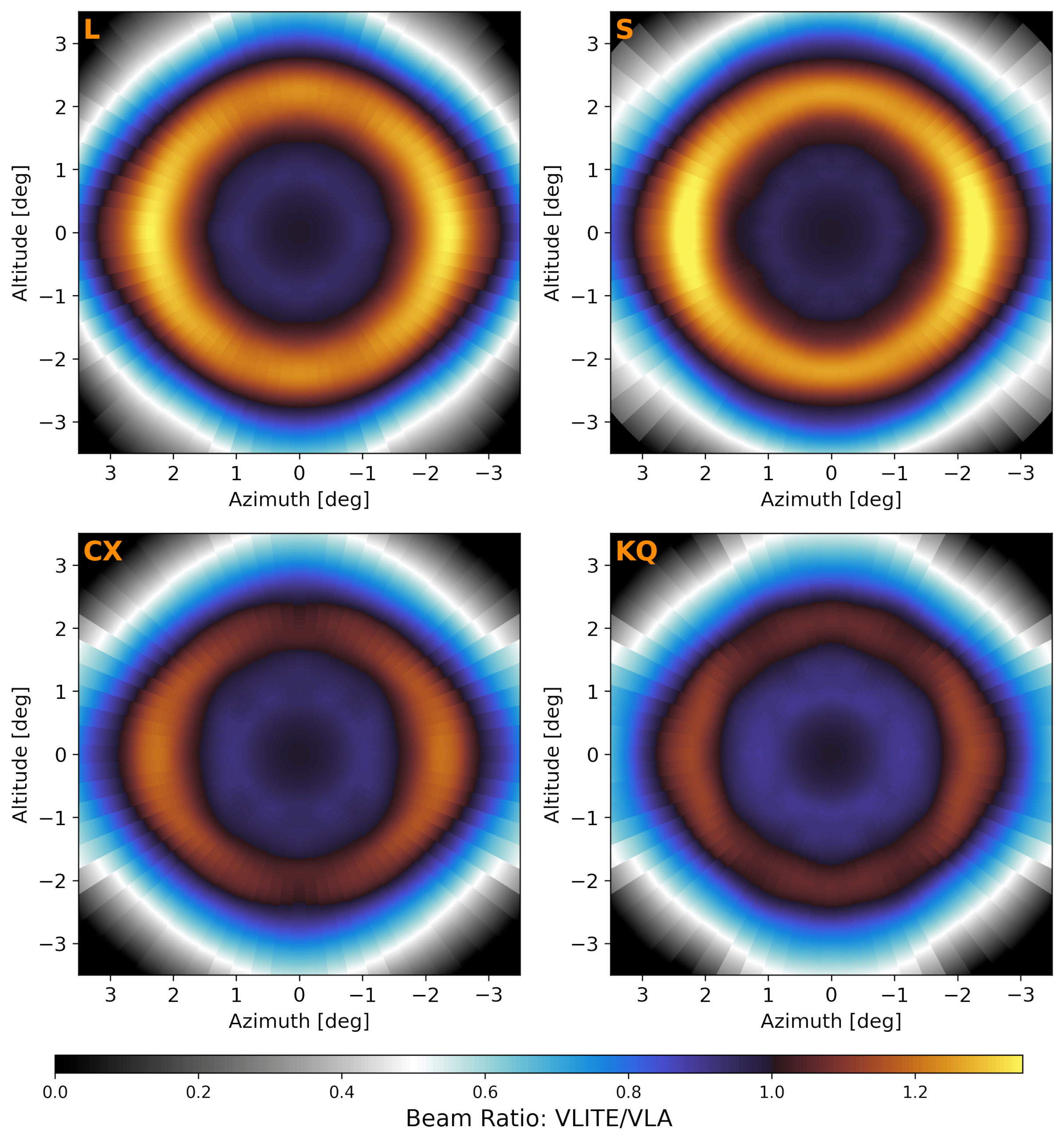}
\end{center}
\caption{Illustration of the ratio of VLITE to VLA primary beam models. In the main lobe ($0.5^\circ - 1.5^\circ$), VLITE models are narrower, resulting in 5-10\% lower power compared to the VLA model. Conversely, in the outer region ($1.8^\circ - 2.5^\circ$), VLITE models exhibit 5-35\% greater plateau strength than the VLA model due to the smaller separation between the subreflector and dipoles, particularly in the L and S beam models. Additionally, VLITE models show asymmetries caused by the tilted subreflector and support struts.}\label{fig:perley}
\end{figure}

We use the polynomial coefficients from the 344 MHz holographic model for comparison with our 340 MHz VLITE models. We extend the polynomial beyond $2.6^\circ$ using linear extrapolation to facilitate this comparison. Figure \ref{fig:perley} shows the ratio of VLITE models to the VLA model. The most striking feature is the difference in the strength of the outer beam plateau: it becomes more pronounced as the subreflector moves away from the prime focus and closer to the dipoles. The CX and KQ models, which correspond to subreflector positions closest to the position during dedicated observations, show the least deviation, with the plateau only 5-20\% stronger than the holographic model. In contrast, the plateau in the S and L-band models is 20-35\% stronger than the holographic model due to the smaller distance between the subreflector and dipoles.

The situation is reversed in the inner $\sim 1.5^\circ$. The S and L beam models exhibit about 5\% less power compared to the holographic model, while the CX and KQ models show a 5-10\% decrease. This indicates that the main lobe in the VLITE models is narrower, with its width varying with subreflector distance, as previously discussed. However, it is curious the discrepancy increases as the VLITE models approach the subreflector position of the VLA model. Fitting our Gaussian model to the VLA polynomial at $\theta < 1.3^\circ$ yields $\sigma = 1.08^\circ$, compared to $\sigma = 1.02^\circ - 0.99^\circ$ for VLITE, indicating a potential small but systematic difference in the primary beam modeling techniques.

The model ratios decrease rapidly beyond $\sim 2.6^\circ$, which is expected due to the extrapolation of the symmetric VLA model beyond its valid range. The observed downward slope and high asymmetry of the outer plateau, however, are genuine features.

\subsection{Image Calibration Accuracy}\label{sec:calibtests}

We next evaluate the accuracy of our image calibration method calculated with the primary beam models, as described in Section \ref{sec:imagecalib}.

\begin{figure}[h!]
\begin{center}
\includegraphics[width=\linewidth]{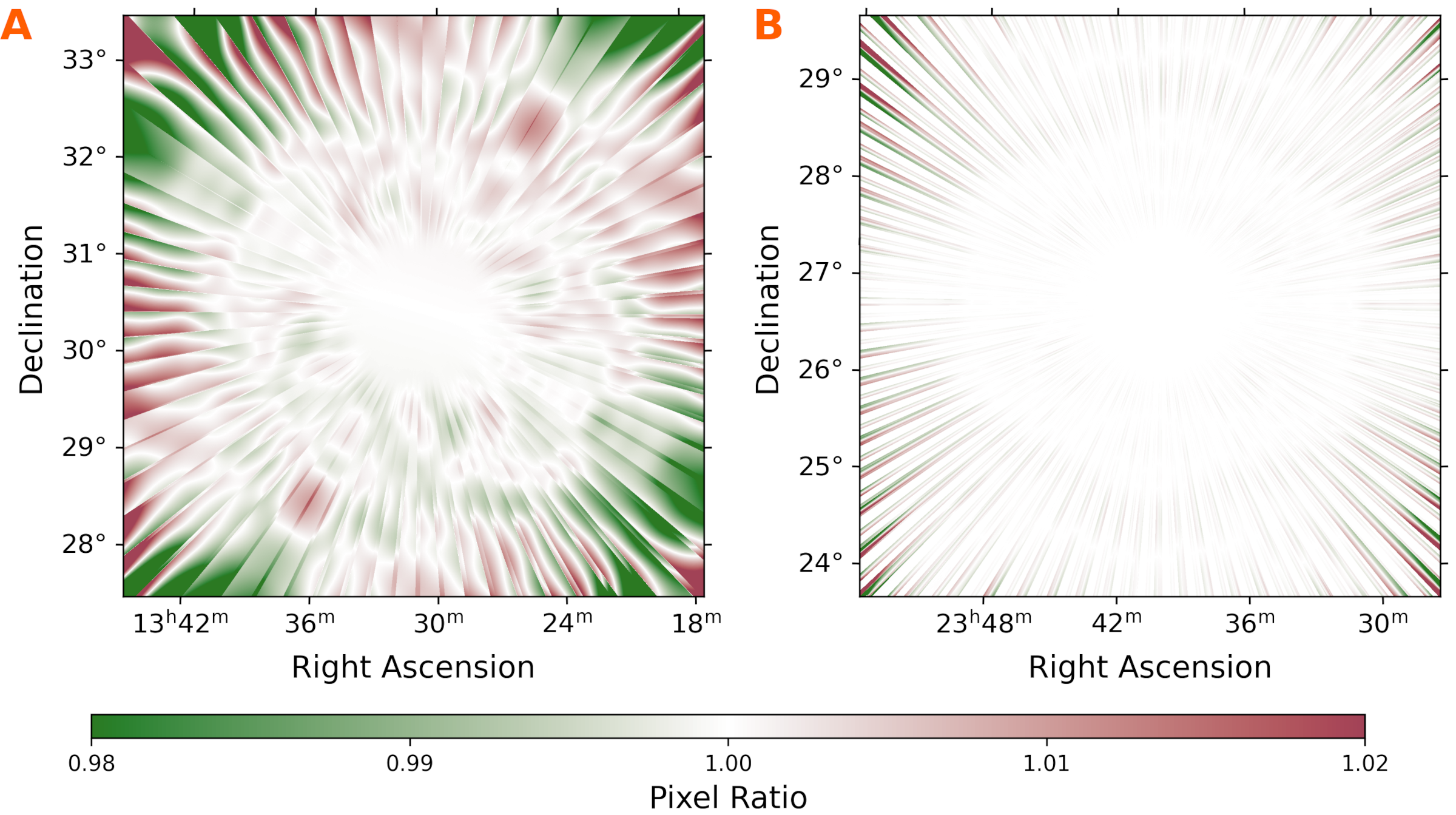}
\end{center}
\caption{Tests of image calibration using discrete sampling of the smeared beam. The primary beam attenuation ratio across the field is calculated using our adopted smear time of 900 s and a fine sampling of 2 s. \textbf{(A)} A 389 s observation of the 3C286 field. The observation was conducted in one continuous block during which the beam rotated by $15^\circ$. \textbf{(B)} A 630 s observation of the J2340+2641 field. The observation consisted of four blocks, with the beam rotation changing by $17^\circ$.} \label{fig:smear}
\end{figure}

First, we assess the impact of using different smear times for image attenuation functions: our chosen smear time of 900 seconds versus a shorter, 2-second smear. The 2-second smear approximates a continuous beam rotation by matching the VLITE data integration time. Our analysis reveals that inaccuracies from using the coarser smear time arise primarily from less precise sampling during long observing blocks, particularly at declinations and hour angles where the parallactic angle changes rapidly. In contrast, long-duration observations with multiple blocks are less affected due to increased sampling frequency. The greatest discrepancies occur in images shorter than 900 seconds and consisting of a single observing block. Figure~\ref{fig:smear} illustrates the ratio of primary beam attenuation using the 900-second smear compared to the 2-second smear.

Figure~\ref{fig:smear}A depicts a $6^\circ$ wide field where the primary beam rotated $15^\circ$ during a continuous 389-second observation. With the 900-second threshold, the smeared beam is approximated with two samples, one at the beginning and one at the end of the observation block. This coarse sampling over a significant rotation angle introduces errors of a few percent, primarily in the image corners. However, at distances $< 3^\circ$, the attenuation function inaccuracies are limited to $2\%$.

Figure~\ref{fig:smear}B shows a similar scenario where the primary beam rotated $17^\circ$ over 630 seconds but the observation was divided into four blocks. The additional samples introduced by the blocks reduce the change in rotation angle between samples, minimizing inaccuracies in the attenuation function to $<1\%$. This is significant because although approximately $40\%$ of VLITE images consist of a single observing block shorter than 900 seconds, fewer than $0.1\%$ of images experience rotation angle changes greater than $10^\circ$. Consequently, we conclude that our discrete sampling of the rotating primary beam is accurate to within $2\%$ for the vast majority of VLITE images.

We next evaluated the accuracy of our primary beam calibration using source light curves. Our beam models were derived from data collected across five C configuration cycles between July 2017 and January 2023. To mitigate potential bias from unequal sampling — where some VLITE fields are observed more frequently than others, potentially skewing results if a small subset of sources is sampled more often and has experienced systematic brightness changes since WENSS observations — we verify the accuracy of our primary beam calibration by assessing sources outside the WENSS footprint. 

We selected bright, unresolved sources from the VLITE C configuration catalog, ensuring each had 15 or more detections and had no matches in the WENSS catalog. Selection criteria included a median S/N $> 30$, a median compactness metric $> 1$, and at least 95\% of detections as single-component Gaussian fits. To ensure comprehensive sampling of the beam attenuation image, we also required that fewer than half of the detections be at the image center. This resulted in 1013 sources with brightnesses ranging $0.1 - 10$ Jy/bm. We then analyzed the 31,895 detections of these sources. 

\begin{figure}[h!]
\begin{center}
\includegraphics[width=\linewidth]{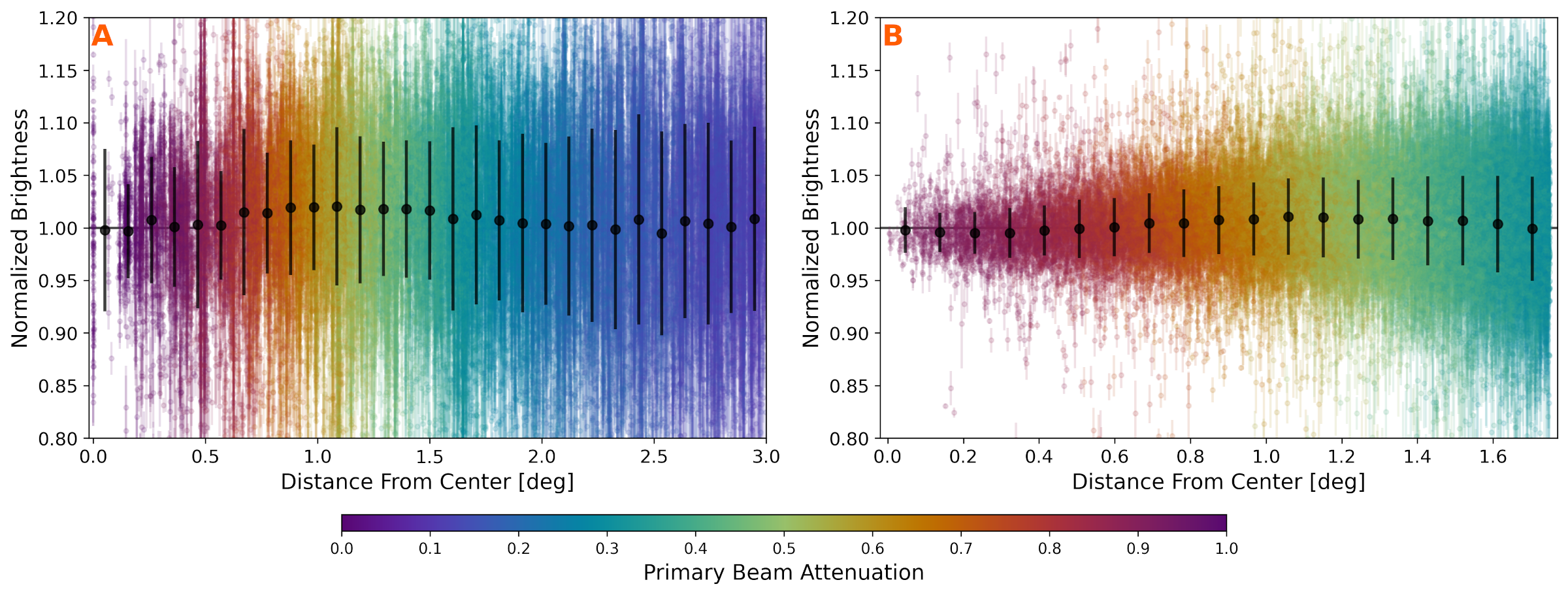}
\end{center}
\caption{Primary beam calibration tests using detections of well-sampled unresolved sources. Each detection is normalized by the average source brightness and plotted according to its distance from the image center. Detections are colored by the primary beam attenuation at their location. \textbf{(A)} C configuration sources outside the WENSS footprint. \textbf{(B)} Bright sources from VCSS epoch 2. Binned averages (black points) show that calibration uncertainty is approximately 1-2\% up to $3^\circ$.}\label{fig:check}
\end{figure}

Figure \ref{fig:check}A shows the calibrated brightness of each detection normalized by the average brightness of the source, plotted by its distance from image center. The detections are colored according to the primary beam attenuation value at each source location. The binned averages reveal a slight bias, with values approximately 2\% higher around $0.7-1.5^\circ$ from image center.

Figure~\ref{fig:check}B presents similar data for sources from the VCSS snapshots catalog during the second observing epoch. The high overlap of VCSS snapshots makes them ideal for testing image calibration with the S-band beam model. Bright sources are typically sampled by $40-50$ snapshots at a variety of positions throughout the $3.5^{\circ}$ wide image field of view, due to the narrow $7.2^{\prime}$ declination spacing of the VLASS observing strips.

We analyzed 33,715 detections from 839 sources, each with an average brightness exceeding 1 Jy/bm, more than 10 detections, median compactness $> 1$, and over 90\% single-component Gaussian fits. We required all source detections to occur on the same day to avoid flux calibration uncertainties. These detections also show a slight upward bias of approximately 1\% at similar distances from image center, a trend that persisted across varied source selection criteria and VCSS epochs. 

To investigate whether the spline method used for the outer beam fits contributed to this bias, we recalculated the beam models using the LOWESS algorithm \citep{lowess}. The fitted values were up to $\sim 1\%$ higher at distances $1-2.5^\circ$ from beam center across all four VLITE beams, potentially accounting for the observed calibrated brightness bias. Unlike the spline method, which weights data by the inverse variance of the measurements, LOWESS applies weighted local linear fits based on proximity to the sampling locations. Sources in long-duration images have lower variance and may have biased the spline fit to lower values due to beam smearing. Additionally, the Gaussian model used for the inner beam may also contribute to the bias, as the VLA polynomial model (green line in Figure \ref{fig:fitbeam}A) shows a slight tendency to higher values around $1^\circ$ from beam center.

Overall, this bias is minimal, and we conclude that the image calibration solutions derived from the primary beam models are accurate to $< 3\%$.

\section{Discussion}

In this paper, we presented a novel technique for mapping and calibrating the primary beam response of VLITE, a commensal telescope situated near the prime focus of an offset Cassegrain optical path. We demonstrated that our method for deriving the unique calibration function for each VLITE image achieves total intensity source measurements with an accuracy of 3\% across a 25 square degree solid angle. This calibration advancement allows for the utilization of VLITE's distinctive defocused primary beam characteristics -- specifically, the absence of nulls and the broad plateau surrounding the main lobe -- for astrophysical plasma research.

\begin{figure}[h!]
\begin{center}
\includegraphics[width=\linewidth]{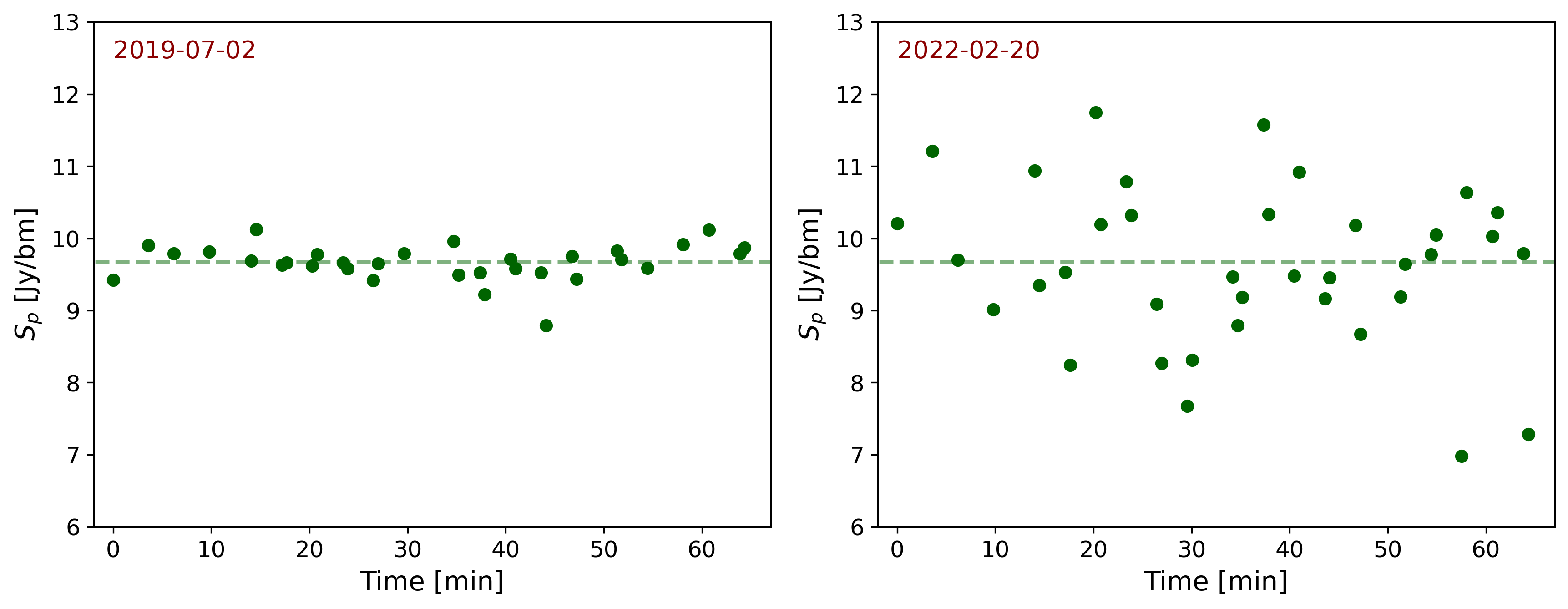}
\end{center}
\caption{Calibrated VCSS light curves for the compact source PKS 2135-20. Left: observation on 2019-07-02, when the source was $140^\circ$ from the Sun, shows a steady light curve. Right: observation on 2022-02-20, when the source was $13^\circ$ from the Sun, reveals strong scintillations. Dashed lines indicate the average brightness from 2019-07-02. Measurement uncertainties are smaller than the plotted markers.}\label{fig:ips}
\end{figure}

One of VLITE's primary scientific objectives is to detect radio transient and variable sources. Achieving accurate measurement of source light curves across a large field of view significantly advances this goal. For example, Figure~\ref{fig:ips} displays the calibrated VLITE light curve of a compact AGN potentially exhibiting interplanetary scintillations (IPS) caused by turbulence in the Solar Wind plasma. IPS enables the measurement of electron density irregularities in the heliosphere \citep{tappin}, which supports space weather research. Moreover, IPS can probe radio sources with sub-arcsecond components without requiring very long baseline interferometry \citep{mwaips}. This capability is valuable for analyzing the size of structures in radio galaxies, aiding in their evolutionary study and potentially providing a new method for identifying high-redshift radio galaxies \citep{mwaips2}.

The opening of the field of view through accurate primary beam calibration has revolutionized the search for transient emissions with VLITE. \cite{pol2023} utilized the initial two VCSS epochs to search for explosive flaring events attributed to the centrifugal breakout of stellar wind plasma trapped in the magnetospheres of early-type stars with $\sim$ kiloGauss strength magnetic fields. They identified three such events with estimated radio luminosities among the highest ever recorded for stellar flares. Although these detections were statistical with a $\sim 1\%$ probability of being false associations, this research highlights VLITE's expanded capability  to discover new transient phenomena and identify promising targets for follow-up observations.

Calibrated VLITE images also open new possibilities for discovering radio transients based on their spectral index. Incorporating spectral information into the selection of transient candidates in radio surveys is a largely unexplored concept, mainly due to the difficulty of obtaining simultaneous measurements across widely separated frequencies. \cite{yuyang2024} conducted a proof-of-concept study by searching for transients with inverted spectra (i.e., brighter at higher frequencies) in epoch 1 of VLASS and VCSS. Inverted spectra typically indicate synchrotron self-absorption in optically thick and dynamically young synchrotron-emitting plasma. Their search identified 22 bright, slow-evolving extragalactic transient candidates unassociated with known AGN. These candidates are likely relativistic tidal disruption events \citep{Somalwar2023J0243} of stars by supermassive black holes (SMBH), though highly variable or transient AGN \citep{Nyland2020} are also possible. This work highlights the powerful technique of using simultaneous survey data to identify new transient radio phenomena.

Primary beam calibration allows for the combination of images to enhance sensitivity. Polisensky et al.\ (in preparation) analyzed 136 hours of observations of the {\it Hubble} Ultra Deep Field (HUDF) collected over 45 days during two consecutive A configuration cycles. By combining a 20-hour subset of high-quality data, they created a deep integration image with a root mean square sensitivity of 158 $\mu$Jy/bm, cataloging 671 sources over 4 square degrees. Among these sources, they identified 6 exhibiting 20-30\% flux density variations over a 1.3-year timescale, with 3 showing MHz-peaked spectra indicative of young AGN plasma jets. 

Nyland et al.\ (in preparation) present Very Long Baseline Array imaging of one of these sources, J0330-2730, revealing a double-lobed morphology with an extent of 140 pc -- over an order of magnitude smaller than predictions based on standard synchrotron self-absorbed jet models. They propose that J0330-2730 is likely a frustrated jet, suggesting that alternative energy loss mechanisms are significant. This research underscores VLITE’s potential for conducting large-scale systematic searches for compact AGN jets, providing new insights into the evolution and life cycles of AGN and the complex interplay between SMBH growth and stellar formation over cosmic time.

Morphological and spectral analysis of extended radio emission in dying and restarted radio galaxies has also been enabled by primary beam calibration \citep{gia2021}. These galaxies exhibit large-scale, extended emission resulting from intermittent jet activity throughout their lifetimes. However, they are relatively rare due to rapid radiative and adiabatic expansion losses that cause their emission to fade within $\sim 10^7$ years. These galaxies are most effectively detected at low frequencies, where their steep spectra, characteristic of aged electron plasma populations, are more pronounced. Calibrated VLITE data thus provides new opportunities to compile larger samples of restarted radio galaxies, offering deeper insights into these systems and their surrounding environments.

The advancement in primary beam calibration has opened new possibilities for pulsar discovery with VLITE. Low-frequency imaging surveys leverage pulsars' compact size and steep radio spectra, offering significant advantages over traditional time-domain surveys. Pulsar detection is favored at low frequencies due to their brightness, but dispersive smearing and scattering at these frequencies can hinder detection by time-domain methods. Imaging surveys, however, are unaffected by these distortions and can detect fast-spinning pulsars in compact or highly eccentric orbits missed by timing surveys. 

Recent efforts have focused on expanding the pulsar census using VLITE's capabilities to survey compact, steep-spectrum sources. An initial search of VLITE images from 97 globular clusters led to the discovery of the first pulsar in cluster GLIMPSE-C01 and marked VLITE's inaugural pulsar discovery \citep{mccarver2024}.

\section*{Conflict of Interest Statement}

The authors declare that the research was conducted in the absence of any commercial or financial relationships that could be construed as a potential conflict of interest.

\section*{Author Contributions}

EP: Conceptualization, Data curation, Formal analysis, Investigation, Methodology, Software, Validation, Visualization, Writing – original draft; TEC: Conceptualization, Funding Acquisition, Writing - review \& editing, Resources ; SG: Conceptualization, Validation, Writing – review \& editing; WP: Data curation, Software, Validation, Writing – review \& editing.

\section*{Funding}
Basic research in radio astronomy at the U.S.\ Naval Research Laboratory is supported by 6.1 Base funding. Construction and installation of VLITE was supported by the NRL Sustainment Restoration and Maintenance fund. The National Radio Astronomy Observatory is a facility of the National Science Foundation operated under cooperative agreement by Associated Universities, Inc. 

\section*{Acknowledgments}

We acknowledge the contribution of Emily Richards to the development of VDP by an appointment to the NRC Research Associateship Program at the US Naval Research Laboratory, administered by the Fellowships Office of the National Academies of Sciences, Engineering, and Medicine. Many student interns supported through the Naval Research Enterprise Internship Program (NREIP) contributed to early efforts to map the VLITE primary beam, including: Lee Bernick, Alice Zhang, and Errol Rochester. Parts of the results in this work make use of the colormaps in the CMasher package \citep{cmasher}.

\bibliographystyle{Frontiers-Harvard} 
\bibliography{vlitepbcor}

\clearpage

\end{document}